\documentclass[aps,pre,showpacs,twocolumn]{revtex4}
\usepackage{bm}
\usepackage{graphicx}
\bibstyle{approve.bib}
\usepackage{amssymb}
\usepackage{amsmath}
\usepackage{esint}
\usepackage{epstopdf}
\usepackage{color}
\usepackage{gensymb}
\usepackage{mathtools}
\usepackage{suffix}

\newcommand{\be}{\begin{equation}}
\newcommand{\ee}{\end{equation}}
\newcommand{\bea}{\begin{eqnarray}}
\newcommand{\eea}{\end{eqnarray}}

\newcommand{\br}{\mathbf{r}}
\newcommand{\bR}{\mathbf{R}}

\newcommand{\ba}{{\boldsymbol{a}}}
\newcommand{\bE}{\mathbf{E}}

\newcommand{\bo}{\mathbf{\Omega}}

\newcommand{\e}{\varepsilon}

\newcommand{\pa}{\parallel}

\newcommand{\ce}{_{\rm c}}

\newcommand{\s}{_{\rm s}}
\newcommand{\kas}{\kappa_{\rm s}}

\newcommand{\hn}{\hat{\rho}}
\newcommand{\hu}{\hat{u}}
\newcommand{\hf}{\hat{F}}

\newcommand{\SB}[1]{\textcolor{black} {#1}}

\begin{document}

\title{Contribution of dipolar bridging to phospholipid membrane interactions: a mean-field analysis}

\author{Sahin Buyukdagli$^{1}$\footnote{email:~\texttt{buyukdagli@fen.bilkent.edu.tr}}  
and Rudolf Podgornik$^{2,3,4}$\footnote{email:~\texttt{podgornikrudolf@ucas.ac.cn \\
{\rm Also affiliated: Department of Physics, Faculty of Mathematics and Physics, University of Ljubljana, SI-1000 Ljubljana, Slovenia.}}}}
\address{$^1$Department of Physics, Bilkent University, Ankara 06800, Turkey\\
$^2$School of Physical Sciences and Kavli Institute for Theoretical Sciences,
University of Chinese Academy of Sciences, Beijing 100049, China\\
$^3$CAS Key Laboratory of Soft Matter Physics, Institute of Physics,
Chinese Academy of Sciences (CAS), Beijing 100190, China\\
$^4$ Wenzhou Institute of the University of Chinese Academy of Sciences, Wenzhou, Zhe-jiang 325000, China.}

\begin{abstract}

We develop a model of interacting zwitterionic membranes 
with rotating surface dipoles 
immersed in a monovalent salt, 
and implement it in a field theoretic formalism. In the 
mean-field regime of monovalent salt, the electrostatic forces 
between the membranes are characterized by a non-uniform trend: at large membrane separations, the interfacial dipoles on the opposing 
sides behave as like-charge cations and give rise to repulsive membrane interactions; 
at short membrane separations, the anionic field induced by the dipolar phosphate groups sets the behavior in the intermembrane region. The attraction of the cationic nitrogens in the dipolar lipid headgroups 
leads to the adhesion of the membrane surfaces via {\sl dipolar bridging}. 
The underlying competition between the opposing field components of the individual dipolar charges leads to the non-uniform salt ion affinity of the zwitterionic membrane with respect to the separation distance; 
large inter-membrane separations 
imply {\sl anionic excess} while 
small, nanometer size separations, favor {\sl cationic excess}. This complex ionic selectivity of zwitterionic membranes may have relevant repercussions on nanofiltration and nanofluidic transport techniques. 
\end{abstract}
\date{\today}
\maketitle   

\section{Introduction}

Macro-molecular forces driven by electrostatics play a primary role in the stability of living matter as well as the regulation of the underlying biological processes~\cite{Isr,PodgornikRev,biomatter}. From the aggregation of like-charged macromolecules~\cite{Molina2014,exp2,Cruz} to the emergence of anionic streaming currents through negatively charged pores~\cite{Heyden2006} and anionic polymer mobility along the applied electric field~\cite{Qiu2015}, electrostatic interactions have been also at the origin of numerous seemingly counterintuitive phenomena. Among these unconventional electrostatic effects, the repulsion between overall neutral dipolar membranes has been identified first in Parsegian's experiments with egg lecithin bilayers as {\sl hydration interaction}~\cite{Parsegian1979,Rand1981}. Subsequently, these experiments have been extended to understand the effect of monovalent salt~\cite{McIntosh1990, Parsegian1991, Kekicheff2014}, divalent buffers \cite{Petrache2011}, and multivalent ions \cite{Averbakh2000,Fink2020} on zwitterionic membrane interactions.

Early theoretical analyses proposed the interactions mediated by perturbed structural order of water molecules~\cite{Marcelja} leading to repulsive hydration interactions~\cite{Leikin}.  Later, the underlying mechanism behind these repulsive hydration forces was surmized to be the image dipoles associated with the interfacial zwitterionic charges~\cite{Jonsson1983,Kjellander1984}. The structural solvation interaction was further detailed in an effective solvent model based on a non-local dielectric response formalism that explained the dipolar membrane repulsion in terms of the coupling between the solvent and surface dipoles~\cite{Belaya1,Belaya2}. \SB{A thermodynamic model of hydration interactions was formulated for complex lipid bilayers with dipolar headgroups based on several spatially separated charges and internal conformational degrees of freedom in Ref.~\cite{Bohinc2017}.} Finally, recent solvent explicit simulations indicated that although the direct electrostatic interaction of the dipolar charges on the opposing membrane walls is attractive, this attraction is reversed by water-mediated hydration forces contributing the major part to the zwitterionic membrane repulsion~\cite{sim1,sim2,sim3}. These studies also showed that various additional effects such as the entropy of the interacting lipid headgroups associated with their annealed degrees of freedom play a non-negligible role in the inter-membrane coupling. 

Due to the complexity of the interaction picture emerging from the simulation results, numerical studies have to be completed by simpler theoretical formalisms including the essential details of the physical system. Motivated by this point, we develop a new model of zwitterionic membranes and implement it in a field theoretic formalism to study the electrostatic mean-field (MF) interaction regime in monovalent salt bathing solutions, where the model provides us with analytically transparent predictions.

We introduce a {\sl dipolar membrane model} in Sec.~\ref{fm}, composed of an electrolyte confined by two membrane walls coated by rotating surface dipoles. By including this orientational degree of freedom, the present formalism generalizes our earlier work on fixed dipolar membrane interactions to realistic zwitterionic charges~\cite{DipFix}. In the model, the electrolyte ions in chemical equilibrium with the external bulk reservoir are treated in the {\sl grand-canonical ensemble}, while the surface dipoles of fixed number are incorporated within the {\sl canonical ensemble}. We then derive the saddle-point equation within the field theoretic formalism and establish the corresponding form of the contact value theorem, that provides us with the MF-level disjoining pressure. Based upon this formalism we characterize in Sec.~\ref{res} the interaction of zwitterionic membranes in MF conditions of monovalent salt and reveal a membrane complexation mechanism driven by the competition between large separation repulsive membrane interactions and short separation zwitterionic charge attractions mediated by {\sl dipolar bridging} interactions. Our main results are summarized and potential extensions of the zwitterionic membrane model are discussed in Conclusions.

\section{Theory and Model}
\label{fm}

\subsection{Field theoretic model}

\begin{figure}
\includegraphics[width=1.\linewidth]{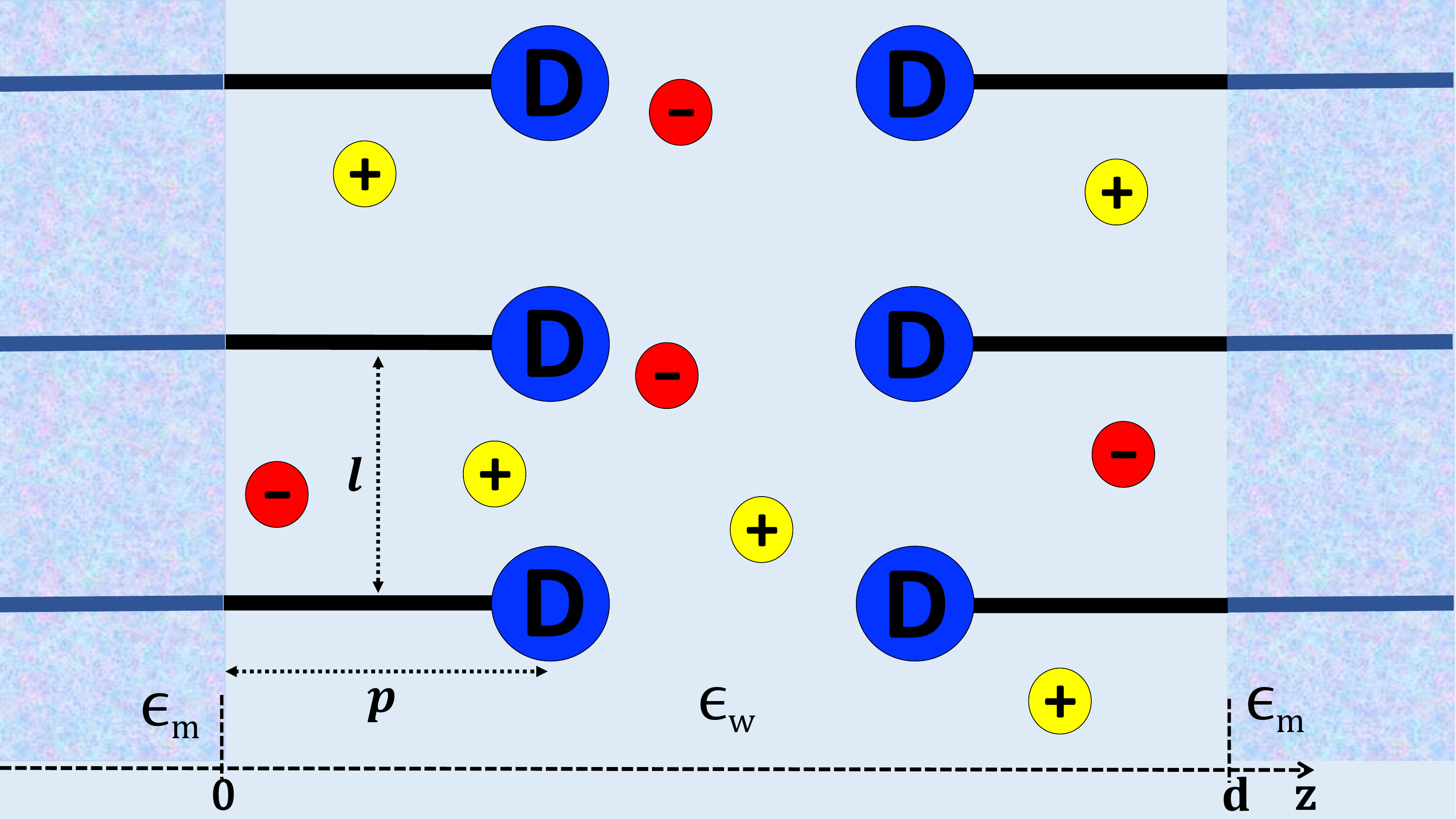}
\caption{(Color online) Side view of the lipid membrane model including the slit pore between two membranes at separation $d$. The surface dipoles (D) are separated by the lateral distance $l$, and located at the distance $p$ from the membrane surface of total area $S$. The membrane and electrolyte permittivities are $\e_{\rm m}$ and $\e_{\rm w}$, respectively.}
\label{fig1}
\end{figure}

In this part, we introduce our dipolar membrane model depicted in Fig.~\ref{fig1}. The two membrane surfaces are located in the $x-y$ plane and are separated by the distance $d$ along the $z$ axis. The surfaces connected to an ion reservoir confine an electrolyte mixture at temperature $T=300$ K. The electrolyte is composed of $s$ ionic species. The ions of the species $n$ with total number $N_n$ have valency $q_n$, fugacity $\lambda_n$, and reservoir concentration $\rho_{{\rm b}n}$. The system is characterized by the dielectric permittivity profile 
\be\label{diel}
\e(\br)=\e_{\rm m}\left[\theta(-z)+\theta(z-d)\right]+\e_{\rm w}\theta(z)\theta(d-z),
\ee
where $\e_{\rm m}$ and $\e_{\rm w}=80$ stand for the dielectric permittivity of the membrane and solvent, respectively.

\begin{figure}
\includegraphics[width=1.\linewidth]{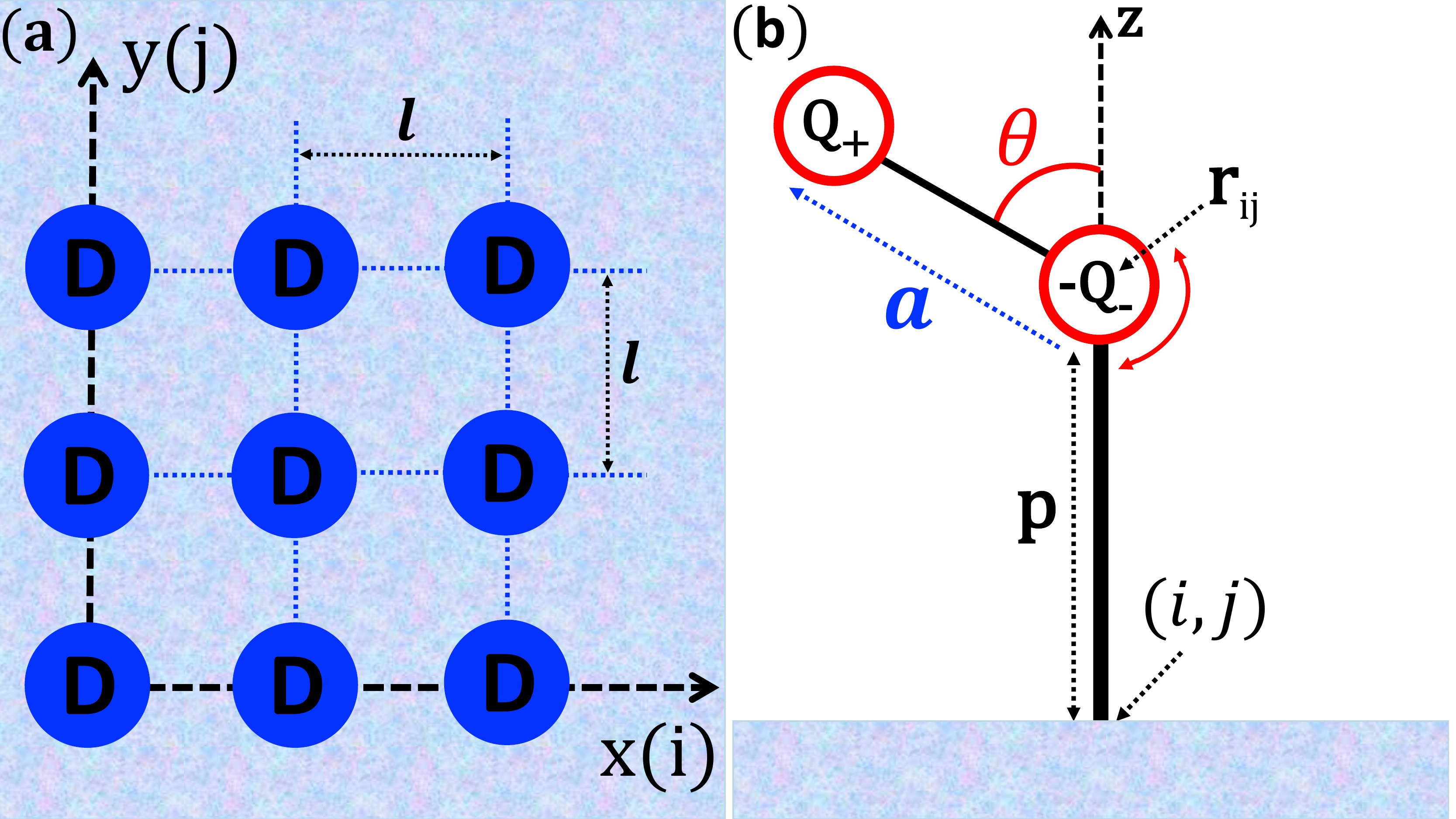}
\caption{(Color online) (a) Lateral view of the lipid membrane surfaces lying in the $x-y$ plane, and located at $z=0$ and $z=d$. The surface dipoles (D) separated by the distance $l$ are located at the discrete mesh points numbered by the indices $i$ and $j$ along the $x$ and $y$ axis, respectively. (b) The inner charge structure of the surface dipoles of length $a$, with an annealed coordinate of the cation that is allowed to rotate about the anionic charge fixed at $z=p$, with $a\leq p$ and $d\geq2p+a$.}
\label{fig2}
\end{figure}

Each membrane wall of total area $S$ carries surface dipoles (D). The charge structure of each dipole is described by an anion ($-Q_-$), fixed at the separation distance $p$ from the wall, and a cation ($Q_+$) at a distance $a$ from the anion,  free to rotate around the anion at a fixed $a$. The wall distribution of the dipoles is illustrated in Fig.~\ref{fig2}(a). The dipoles separated from each other by the lateral distance $l$ are located at the discrete mesh points of square symmetry, numbered by the indices $i$ and $j$ along the $x$ and $y$ axis, respectively. Each row or column on the membrane surfaces contains a total of $N$ dipoles, i.e. $1\leq i\leq N$ and $1\leq j\leq N$. Thus, the position vectors $\br_{ij}$ and $\br'_{ij}$ of the dipoles respectively located on the left and right surfaces and characterized by the solid angles $\bo_{ij}$ and $\bo'_{ij}$ read
\bea
\label{pos1}
\br_{ij}&=&(i-1)l\hu_x+(j-1)l\hu_y+p\hu_z,\\
\label{pos2}
\br'_{ij}&=&(i-1)l\hu_x+(j-1)l\hu_y+(d-p)\hu_z.
\eea

The total charge density operator is
\bea
\label{ch1}
\hn\ce(\br)&=&\sum_{n=1}^sq_n\sum_{m=1}^{N_n}\delta(\br-\bR_{nm})\\
&&+\sum_{i=1}^N\sum_{j=1}^N\left[\hf\left(\br;\br_{ij},\bo_{ij}\right)+\hf\left(\br;\br'_{ij},\bo'_{ij}\right)\right],\nonumber
\eea
where $\bR_{nm}$ stands for the mobile ion coordinate, and $\hf\left(\br;\br_{ij},\bo_{ij}\right)$ is the charge structure factor of the dipole $(i,j)$. The canonical partition function of the system reads
\be
\label{can1}
Z\ce=\prod_{n=1}^s\prod_{m=1}^{N_n}\int\mathrm{d}^3\bR_{nm}\prod_{i=1}^N\prod_{j=1}^N\int\SB{\frac{\mathrm{d}^2\bo_{ij}}{4\pi}\frac{\mathrm{d}^2\bo'_{ij}}{4\pi}}e^{-\beta E\ce-\beta E_{\rm n}},
\ee
where the electrostatic interaction energy is given by
\be\label{can2}
\beta E\ce=\frac{1}{2}\int\mathrm{d^3}\br\mathrm{d^3}\br'\rho\ce(\br)v\ce(\br,\br')\rho\ce(\br'),
\ee
In Eq.~(\ref{can2}), the Coulomb Green's function $v\ce(\br,\br')$ is defined in terms of its inverse, $v\ce^{-1}(\br,\br')=-(k_{\rm B}T/e^2)\nabla\cdot\e(\br)\nabla\delta(\br-\br')$. Moreover, In Eq.~(\ref{can1}), the total steric energy of the ions and dipoles is defined as
\bea
\label{can3}
\beta E_{\rm n}&=&\sum_{n=1}^s\sum_{m=1}^{N_n}V_{\rm n}(\bR_{nm})\\
&&+\sum_{i=1}^N\sum_{j=1}^N\left[V_{\rm l}(\br_{ij},\bo_{ij})+V_{\rm r}(\br'_{ij},\bo'_{ij})\right],\nonumber
\eea
where we introduced the ionic onsite potential $V_{\rm n}(\bR_{nm})$, and the dipolar onsite potentials $V_{\rm l}(\br_{ij},\bo_{ij})$ and $V_{\rm r}(\br'_{ij},\bo'_{ij})$, with the indices ${\rm l}$ and ${\rm r}$ referring to the dipoles located on the left and right walls, respectively. \SB{These one-body potentials will be used to generate the ionic charge densities and the conditional probabilities of the surface dipoles.} 

At this point, performing an Hubbard-Stratonovich transformation, we introduce the fluctuating electrostatic potential \SB{$\psi(\br)$} that allows us to linearize Eqs.~(\ref{can1})-(\ref{can2}) in the charge density~(\ref{ch1}). Then, via the replacement $\sum_{i=1}^N\sum_{j=1}^N\to\sigma\s\int\mathrm{d}^2\br_\pa$, we pass to the continuously distributed dipole limit, with the position vector $\br_\pa=x\hu_x+y\hu_y$ within the membrane surface, and the surface dipole density $\sigma\s=l^{-2}$. Finally, taking into account the ion exchange driven by the chemical equilibrium between the slit region and the reservoir, the grand-canonical partition function of the system takes the functional integral form 
\be
\label{can5}
Z_{\rm G}=\prod_{n=1}^s\sum_{N_n=1}^\infty\frac{\lambda_n^{N_n}}{N_n!}Z\ce=\int\mathcal{D}\SB{\psi}\;e^{-\beta H[\SB{\psi}]},
\ee
with the Hamiltonian functional
\bea
\label{can7}
\beta H[\SB{\psi}]&=&\frac{k_{\rm B}T}{2e^2}\int\mathrm{d}^3\br\;\e(\br)\left[\nabla\SB{\psi}(\br)\right]^2\\
&&-\sum_{n=1}^s\lambda_n\int\mathrm{d}^3\br e^{iq_n\SB{\psi}(\br)-V_n(\br)}\nonumber\\
&&-\sigma\s\int\mathrm{d}^3\br\left[\delta(z-p)+\delta(z-d+p)\right]\nonumber\\
&&\hspace{7mm}\times\ln\int\frac{\mathrm{d}^2\bo}{4\pi}e^{i\int\mathrm{d}^3\br'\hf\left(\br';\br,\bo\right)\SB{\psi}(\br')-V_{\rm d}(\br,\bo)}.\nonumber
\eea
In Eq.~(\ref{can7}), the dipolar potential is defined as $V_{\rm d}(\br,\bo)=V_{\rm l}(\br,\bo)$ if $z=p$ and $V_{\rm d}(\br,\bo)=V_{\rm r}(\br,\bo)$ if $z=d-p$. 

\subsection{Derivation of the saddle-point equations}

Fig.~\ref{fig2}(b) displays the inner charge structure of the rotating membrane dipoles with total length $a$ and terminal charges $\pm Q_\pm$. The cationic charge $Q_+$ stands for the valency of the nitrogen (${\rm N}^+$) located at the trimethyl ammonium moiety of the lipid headgroup, and the fixed anionic charge $-Q_-$ is the valency of the phosphate group. The charge structure factor of the surface dipole rotating about the anionic charge located at $\br$ reads 
\be
\label{can13}
\hf(\br';\br,\bo)=-Q_-\delta(\br'-\br)+Q_+\delta(\br'-\br-\ba).
\ee
In order to prevent the rotating cations from touching the membrane surface and the tail of the dipoles on the opposing membrane surface, we will restrict ourselves to the tail length and pore size regimes of $p\geq a$ and $d\geq2p+a$. \SB{Moreover, with the aim to prohibit the contact of the neighbouring dipoles located on the same wall, we will limit our study to the dipolar distance regime of $l>2a$. For the characteristic dipole size $a=2$ {\AA} considered in our work, this assumption sets for the surface dipole density the reasonable constraint of $\sigma_{\rm s}\lesssim 6$ ${\rm nm}^{-2}$.}

Within the MF approximation, the ion density follows from the thermodynamic relation $\rho_n(\br)=-\delta\ln Z_{\rm G}/\delta V_n(\br)\approx\delta H/\delta V_n(\br)$ as $\rho_n(\br)=\lambda_ne^{iq_n\SB{\psi}(\br)-V_n(\br)}$. In the bulk reservoir where $\SB{\psi}(\br)=V_n(\br)=0$, this relates the fugacity to the bulk concentration as $\lambda_n=\rho_{\rm nb}$. Passing to the real \SB{dimensionless potential $\phi(\br)$ via the transformation $\psi(\br)\to i\phi(\br)$}, and accounting for the planar symmetry $\phi(\br)=\phi(z)$, the average ion density becomes
\be\label{can8}
\rho_n(z)=\rho_{n{\rm b}}e^{-q_n\phi(z)-V_n(z)}.
\ee

In order to simplify the notation, from now on, the ionic and dipolar steric potentials will be omitted. Plugging now the dipolar structure factor~(\ref{can13}) into Eq.~(\ref{can7}), and evaluating the saddle-point condition $\delta H[\phi]/\delta\phi(z)=0$, after some algebra, one obtains the zwitterionic Poisson-Boltzmann (ZPB) Eq. in the form
\bea
\label{can14}
\hspace{-5mm}&&\frac{k_{\rm B}T}{e^2}\partial_z\SB{\left[\e(z)\partial_z\phi(z)\right]}+\sum_{n=1}^sq_n\rho_n(z)+Q_+\left[\rho^+_{\rm l}(z)+\rho^+_{\rm r}(z)\right]\nonumber\\
\hspace{-5mm}&&=\sigma\s Q_-\left[\delta(z-p)+\delta(z-d+p)\right].
\eea
In Eq.~(\ref{can14}), we introduced the dipolar cation densities
\be\label{pr2}
\rho^+_{\rm l,r}(z)=\frac{\sigma\s}{2a I_{\rm l,r}}e^{-Q_+\phi(z)}\theta_{\rm l,r}(z)
\ee
derived in Appendix~\ref{ap1}. The auxiliary functions 
\bea
\label{tl1}
\theta_{\rm l}(z)&=&\theta(z-p+a)\theta(a-z+p),\\
\label{tl2}
\theta_{\rm r}(z)&=&\theta(z-d+p+a)\theta(d-p+a-z),
\eea
including the Heaviside step function $\theta(z)$ account for the finite extension of the interfacial layer scanned by the rotating dipoles. Then, the integrals incorporating the constraint on the fixed lateral dipole density read
\be
\label{can15}
I_{\rm l}=\int_{-a}^{a}\frac{\mathrm{d}a_z}{2a}e^{-Q_+\phi(p+a_z)};\;I_{\rm r}=\int_{-a}^{a}\frac{\mathrm{d}a_z}{2a}e^{-Q_+\phi(d-p+a_z)},
\ee
with the dipole projection $a_z=a\cos\theta$. Using the mirror symmetry with respect to the mid-pore, one can show that these integrals are equal, $I_{\rm l}=I_{\rm r}$. Therefore, when the specification of the membrane location is not necessary, these integrals will be denoted by the symbol $I$. 

In Appendix~\ref{ap1}, we show that the ZPB Eq.~(\ref{can14}) can be expressed as a generalized Poisson identity. The iterative solution of the ZPB Eq. can be carried out by setting first the integral~(\ref{can15}) to $I=1$, and updating the value of $I$ at each iterative step until numerical convergence is achieved. This solution scheme should be applied with the electrostatic boundary conditions
\be\label{bc1}
\phi'(z\ce^+)-\phi'(z\ce^-)=2/\mu_-;\hspace{5mm}\phi'(z\s)=0.
\ee
In Eq.~(\ref{bc1}), $z\ce=\{p,d-p\}$ is the position of the fixed dipolar anions. Moreover, the anionic and cationic Gouy-Chapman (GC) lengths are defined as~\cite{Gouy,Chapman}
\be
\label{gc}
\mu_\pm=\left(2\pi Q_\pm\ell_{\rm B}\sigma\s\right)^{-1}.
\ee
Finally, the boundaries of the finite field region read $z\s=\{p-a,d-p+a\}$ for the salt-free liquid, and $z\s=\{0,d\}$ in the presence of added salt.

\begin{figure*}
\includegraphics[width=1.\linewidth]{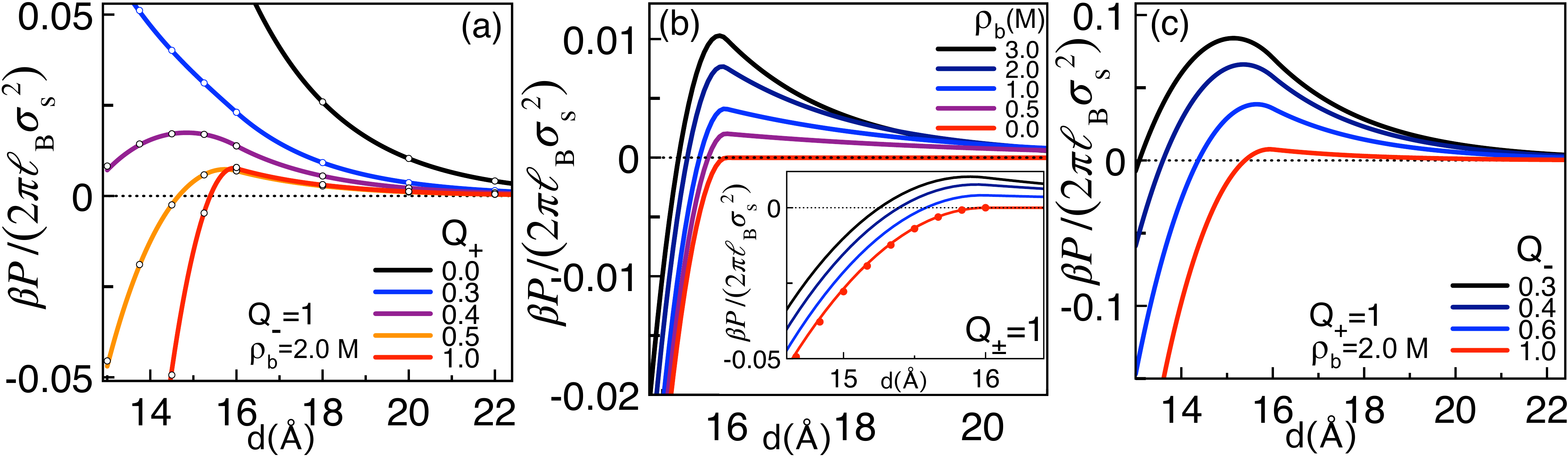}
\caption{(Color online) Interaction pressure between the dipolar membrane walls as a function of the membrane separation  $d$ for varying (a) dipolar cation valency, (b) salt concentration, and (c) dipolar anion valency. Solid lines are from Eq.~(\ref{pr1}) with the numerical solution of the ZPB Eq.~(\ref{can14}), the open circles in (a) display the Debye-H\"{u}ckle-level pressure~(\ref{prc4}), and the red circles in the inset of (b) correspond to Eq.~(\ref{prc2}) for pure solvent. The model parameters are $p=5$ {\AA}, $a=3$ {\AA}, and $\sigma\s=0.1$ $\SB{\rm nm^{-2}}$.}
\label{fig3}
\end{figure*}

\subsection{Contact value theorem}
\label{cn}

Here, we calculate the inter-membrane disjoining pressure and derive the contact value theorem. In the remainder of this work, we consider a symmetric monovalent salt with ionic valencies $q_\pm=\pm1$ and bulk concentrations $\rho_{\pm{\rm b}}=\rho_{\rm b}$. The MF grand potential of the system corresponds to the Hamiltonian~(\ref{can7}) evaluated with the solution of the saddle-point Eq.~(\ref{can14}), 
\bea
\label{fr}
\frac{\beta \Omega_{\rm mf}}{S}&=&-\int_0^d\frac{\mathrm{d}z}{8\pi\ell_{\rm B}}\left[\phi'(z)\right]^2-\int_0^d\mathrm{d}z\left[\rho_+(z)+\rho_-(z)\right]\nonumber\\
&&-\sigma\s Q_-\left[\phi(p)+\phi(d-p)\right]-\sigma\s\ln \left[I_{\rm l}I_{\rm r}\right].
\eea
The MF pressure is obtained from the grand potential~(\ref{fr}) via the identity $P_{\rm in}=-\delta\Omega_{\rm mf}/\delta(Sd)$. Subtracting from the latter the bulk pressure $\beta P_{\rm out}=2\rho_{\rm b}$, after some algebra, the net pressure $P=P_{\rm in}-P_{\rm out}$ follows as
\bea
\label{pr1}
\beta P&=&\frac{\sigma\s Q_-}{2}\left\{\phi'\left[(d-p)^+\right]+\phi'\left[(d-p)^-\right]\right\}\\
&&+\rho_+(d)+\rho_-(d)-2\rho_{\rm b}\nonumber\\
&&+\rho^+_{\rm r}(d-p+a)-\rho^+_{\rm r}(d-p-a).\nonumber
\eea

Eq.~(\ref{pr1}) generalizes the contact value theorem, originally introduced for charged membranes~\cite{Hen1,Hen2,Isr}, to membranes decorated with a dipolar charge. The first line on the r.h.s. of Eq.~(\ref{pr1}) corresponds to the electrostatic pressure, while the second line is the standard osmotic pressure associated with the contact ion density at the membrane surface. Finally, the third line is the osmotic pressure contribution stemming from the surface dipoles. Interestingly, one sees that the dipolar pressure component  involves the density contrast of the surface dipoles between the orientational configurations $\theta=\pi$ and $\theta=0$. \SB{The discrepancy between this form and the standard ion contact density stems from the fact that the dipolar cations, unable to touch the membrane surface, transmit to the latter the mechanical stress originating from their rotational kinetic energy via their tail.}

In the standard PB framework the contact theorem and the first integral of the PB equation are connected. We now analyze their relation in the present model. In Appendix~\ref{apcv}, we derive the first integral of the ZPB Eq.~(\ref{can14}) and obtain the following more general expression 
\bea
\label{cn1}
\beta P&=&-\frac{\phi'(z)^2}{8\pi\ell_{\rm B}}+\sum_{n=1}^s\left[\rho_n(z)-\rho_{n{\rm b}}\right]+\rho^+_{\rm r}(z)+\rho^+_{\rm l}(z)\nonumber\\
&&+\rho^+_{\rm r}(d-p+a)\theta(z-d+p-a)\nonumber\\
&&-\rho^+_{\rm r}(d-p-a)\theta(z-d+p+a)\nonumber\\
&&+\rho^+_{\rm l}(p-a)\theta(p-a-z)-\rho^+_{\rm l}(p+a)\theta(p+a-z)\nonumber\\
&&+\frac{\sigma\s Q_-}{2}\left\{\phi'\left[(d-p)^+\right]+\phi'\left[(d-p)^-\right]\right\}\nonumber\\
&&\hspace{1.25cm}\times\theta(z-d+p)\nonumber\\
&&-\frac{\sigma\s Q_-}{2}\left[\phi'(p^+)+\phi'(p^-)\right]\theta(p-z),
\eea
for the disjoining pressure. In agreement with the mechanical equilibrium constraint, Eq.~(\ref{cn1}) corresponding to the first integral, is of course independent of the coordinate $z$. 

While on first sight the first integral differs from the contact value theorem, it is straightforward to verify that at the point $z=d$, Eq.~(\ref{cn1}) yields consistently exactly the contact value pressure~(\ref{pr1}), derived previously from the variation of the thermodynamic grand potential~(\ref{fr}). Then, at the mid pore, $z=d/2$, one obtains from Eq.~(\ref{cn1}) the following alternative form of the pressure
\bea
\label{cn2}
\beta P&=&\sum_{n=1}^s\left[\rho_n(d/2)-\rho_{n{\rm b}}\right]\\
&&+\left[\rho_{{\rm l}+}(d/2)-\rho_{{\rm l}+}(p+a)\right.\nonumber\\
&&\hspace{3mm}\left.+\rho_{{\rm r}+}(d/2)-\rho_{{\rm r}+}(d-p-a)\right]\theta(p+a-d/2).\nonumber
\eea
One sees that at large separation distances $d>2(p+a)$ where the dipolar terms in Eq.~(\ref{cn2}) vanish, the membrane interactions are purely due to the osmotic pressure difference of the salt ions in the mid-pore and the bulk reservoir. Then, in the close distance regime $d\leq 2(p+a)$ where the dipolar layers on the opposing membranes overlap, the interaction pressure is augmented by the dipolar density contrast between the mid-pore and the perpendicular dipole configuration $\theta=0$. Below, we probe the consequences of this dipolar interaction mechanism on the disjoining pressure of the zwitterionic membranes.

\section{Results}
\label{res} 

Fig.~\ref{fig3}(a) illustrates the effect of freely rotating surface dipoles on membrane interactions. The solid curves show the separation dependence of the disjoining pressure~(\ref{pr1}) computed from the numerical solution of the ZPB Eq.~(\ref{can14}) for different values of the dipolar cation valency $Q_+$. One sees that in the standard case of purely monopolar charged membranes (black curve for $Q_+=0$), the approach of the membranes leads to the monotonous rise of the uniformly repulsive pressure ($d\downarrow P\uparrow$) according to the DH law~\cite{Isr}
\be
\label{prdh}
\beta P\approx2\pi\ell_{\rm B}\sigma\s^2Q_-^2\frac{\cosh^2\left(\kas p\right)}{\sinh^2\left(\kas d/2\right)}
\ee
(see the open circles). Then, as the dipolar cation valency rises towards the zwitterionic dipole limit $Q_+=Q_-$, the short distance branch of the pressure gradually drops and finally turns from repulsive to strongly attractive. Hence, the interaction of neutral surface dipoles gives rise to non-monotonic inter-membrane forces, characterized by a large-separation repulsive regime and a short distance attractive branch, separated by a pressure barrier. This membrane adhesion effect occurring on the level of MF electrostatic conditions is the key prediction of our work.

Fig.~{\ref{fig3}(b) illustrates the influence of added salt on the interaction pressure of a purely zwitterionic membrane with $Q_\pm=1$. One notes that the increment of salt moderately weakens the short-range attractive interactions but strongly amplifies the repulsive pressure branch. Indeed, one sees that in the salt-free case (red curve at $\rho_{\rm b}=0$), the repulsive branch dies out and the intermembrane interactions become purely attractive. This suggests that in zwitterionic membranes, the inter-membrane repulsions occurring at large separations are mediated by salt ions. In order to gain a better physical insight into the features revealed so far, we consider below the analytically tractable limits of the ZPB Eq.~(\ref{can14}).

\subsection{Explaining the membrane adhesion mechanism}

In order to explain the mechanism driving the membrane attraction regime, we consider here the salt-free case ($\rho_{n{\rm b}}=0$), where the ZPB Eq.~(\ref{can14}) possesses an exact analytical solution. The derivation of this solution is detailed in Appendix~\ref{ap2}. As the interfacial dipole layers start to overlap at $d\leq2(p+a)$, the solution of Eq.~(\ref{can14}) has to be considered in two distinct regimes. 

\subsubsection{Decoupled dipole regime $d\geq2(p+a)$}
\label{pa1}

In the separation regime $d\geq2(p+a)$ where the rotating dipoles on different membrane surfaces do not overlap, the solution of Eq.~(\ref{can14}) has the piecewise form
\bea
\label{ppc}
\phi(z)&=&\phi_0\theta(z)\theta(p-a-z)+\phi_1(z)\theta(z-p+a)\theta(p-z)\nonumber\\
&&+\phi_2(z)\theta(z-p)\theta(p+a-z)\nonumber\\
&&+\phi_0\theta(z-p-a)\theta(d/2-z)\\
&&+\phi(d-z)\theta(z-d/2)\theta(d-z),\nonumber
\eea
In Eq.~(\ref{ppc}), $\phi_0$ is a constant reference potential, the non-uniform potential components
\bea
\label{c2}
&&\phi_1(z)=\phi_0+\frac{1}{Q_+}\ln\left\{\cos^2\left[\frac{\kappa_0}{\sqrt 2}(z-p+a)\right]\right\},\\
\label{c2II}
&&\phi_2(z)=\phi_0+\frac{1}{Q_+}\ln\left\{\cos^2\left[\frac{\kappa_0}{\sqrt 2}(z-p-a)\right]\right\}
\eea
are similar in form to the monopolar membrane potential in counterion-only liquids~\cite{Isr}, and the screening parameter $\kappa_0$ satisfies the transcendental equation
\be\label{ts1}
\frac{\kappa_0 a}{\sqrt2}\tan\left(\frac{\kappa_0 a}{\sqrt2}\right)=\frac{Q_+a}{2\mu_-}.
\ee
\SB{Moreover, the fourth line of Eq.~(\ref{ppc}) indicates the mirror symmetry of the potential solution with respect to the point $z=d/2$, i.e. $\phi(d-z)=\phi(z)$. We finally note that due to the absence of dipolar contact at  $d>2(p+a)$, the electrostatic potential~(\ref{ppc}) does not depend on the membrane separation distance $d$.}

\begin{figure*}
\includegraphics[width=.6\linewidth]{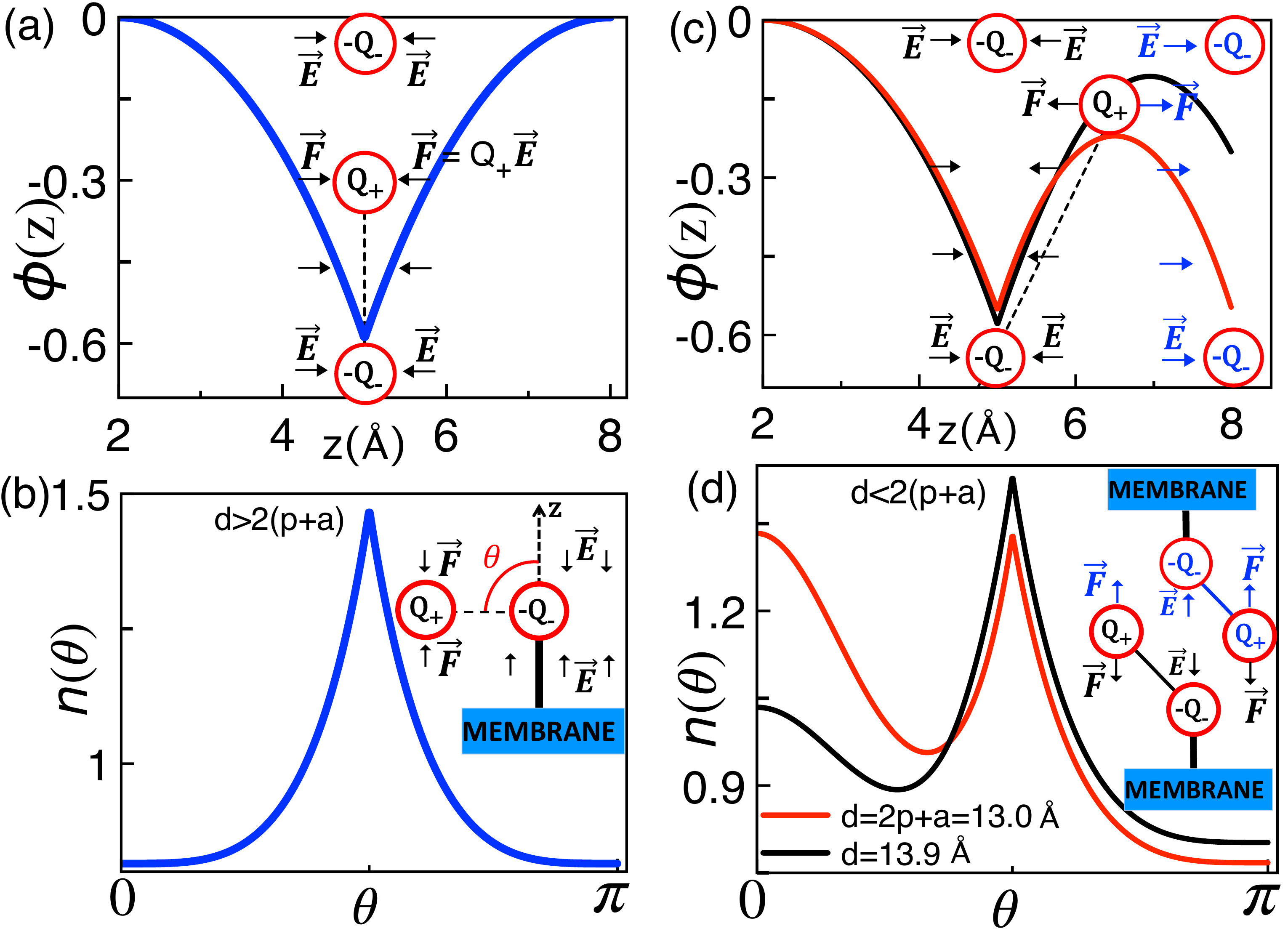}
\caption{(Color online) (a) Electrostatic potential~(\ref{ppc}) and (b) orientational probability~(\ref{or1}) in the large distance regime $d\geq2(p+a)$ of the salt-free system ($\rho_{\rm b}=0$), where the surface dipoles are electrostatically decoupled. (c)-(d) The plots in (a)-(b) in the short separation regime $d<2(p+a)$, where the dipolar layers interact. The black and blue arrows indicate the electric field lines $\bE=-\nabla\phi(z)$ and the resulting force $\mathbf{F}=Q_+\bE$ on the dipolar cations generated by the dipolar anions $-Q_-$ at $z=p$ and $z=d-p$, respectively. The model parameters are $p=5$ {\AA}, $a=3$ {\AA}, $Q_\pm=1$, and $\sigma\s=1.0$ \SB{${\rm nm}^{-2}$}.}
\label{fig4}
\end{figure*}

Fig.~{\ref{fig4}(a) displays the potential~(\ref{ppc}) corresponding to the dipolar cation energy (see Eq.(\ref{pr2})) close to the wall at $z=0$. One sees that the electric field lines $\bE=-\nabla\phi(z)$ pointing towards their anionic charge sources are oriented perpendicular to the membrane (black arrows). The resulting force $\mathbf{F}=Q_+\bE$ drives  the dipolar cations towards the potential minimum at $z=p$. The corresponding dipolar orientation can be identified via the orientational probability density
\be
\label{nd3}
n(a_{\rm z})=\frac{1}{I}e^{-Q_+\phi(p+a_z)}
\ee
derived in Appendix~{\ref{ap1II}. Plugging Eq.~(\ref{ppc}) into Eq.~(\ref{nd3}), the probability function follows as
\be\label{or1}
n(a_{\rm z})=\frac{\kappa a/\sqrt{2}}{\tan\left(\kappa a/\sqrt{2}\right)}\sec^2\left[\frac{\kappa}{\sqrt2}\left(|a_{\rm z}|-a\right)\right].
\ee
Fig.~\ref{fig4}(b) shows that the probability~(\ref{or1}) is maximized at the angle $\theta=\pi/2$. This indicates that in the separation regime $d>2(p+a)$, the electrostatic force $\mathbf{F}$ allignes the dipoles parallel with the membrane surface (see the inset).

In order to calculate the membrane interaction force corresponding to this configuration, we substitute the potential solution~(\ref{ppc}) into Eq.~(\ref{pr1}). This yields $P=0$, which confirms that the red pressure curve in Fig.~\ref{fig3}(b) vanishes at $d\geq2(p+a)$. The cancellation of the interaction force at large separation is due to the vanishing of the electrostatic field $E_z(z)=-\partial_z\phi_0=0$ in the inter-membrane region $p+a<z<d-p-a$ separating the dipolar layers. This leads to the electrostatic decoupling of the membranes, and the invariance of their individual electrostatic energy under the variation of their separation.

\subsubsection{Coupled dipole regime $2(p+a)\geq d\geq2p+a$}
\label{pa2}

For pore sizes located in the region $2(p+a)\geq d\geq2p+a$ where the dipolar regions overlap, the piecewise solution of Eq.~(\ref{can14}) reads
\bea
\label{ppc2}
\phi(z)&=&\phi_0\theta(z)\theta(p-a-z)+\phi_1(z)\theta(z-p+a)\theta(p-z)\nonumber\\
&&+\phi_2(z)\theta(z-p)\theta(d-p-a-z)\nonumber\\
&&+\phi_3(z)\theta(z-d+p+a)\theta(d/2-z)\nonumber\\
&&+\phi(d-z)\theta(z-d/2)\theta(d-z).
\eea
In Eq.~(\ref{ppc2}), the potential components are
\bea
\label{c5}
&&\hspace{-3mm}\phi_1(z)=\phi_0+\frac{1}{Q_+}\ln\left\{\cos^2\left[\frac{\kappa_1}{\sqrt2}(z-p+a)\right]\right\},\\
\label{c6}
&&\hspace{-3mm}\phi_2(z)=\phi_0+\frac{1}{Q_+}\ln\left\{\frac{\kappa_1^2}{\kappa_2^2}\cos^2\left[\frac{\kappa_2}{\sqrt2}(z-z_0)\right]\right\},\\
\label{c7}
&&\hspace{-3mm}\phi_3(z)=\phi_0+\frac{1}{Q_+}\ln\left\{\frac{\kappa_1^2}{\kappa_3^2}\cos^2\left[\kappa_3\left(z-\frac{d}{2}\right)\right]\right\},
\eea
and the parameters $\kappa_1$, $\kappa_2$, $\kappa_3$, and $z_0$ satisfy the coupled transcendental Eqs.
\bea\label{t1}
&&\hspace{-3mm}\cos\left[\frac{\kappa_2}{\sqrt2}(p-z_0)\right]=\frac{\kappa_2}{\kappa_1}\cos\left(\frac{\kappa_1a}{\sqrt2}\right),\\
\label{t2}
&&\hspace{-3mm}\tan\left[\frac{\kappa_2}{\sqrt2}(p-z_0)\right]=\frac{\kappa_1}{\kappa_2}\tan\left(\frac{\kappa_1a}{\sqrt2}\right)-\frac{\sqrt2Q_+}{\kappa_2\mu_-},\\
\label{t3}
&&\hspace{-3mm}\cos\left[\frac{\kappa_2}{\sqrt2}(d-p-a-z_0)\right]=\frac{\kappa_2}{\kappa_3}\cos\left[\kappa_3\left(\frac{d}{2}-p-a\right)\right],\nonumber\\
&&\\
\label{t4}
&&\hspace{-3mm}\sin\left[\frac{\kappa_2}{\sqrt2}(d-p-a-z_0)\right]=\sqrt2\sin\left[\kappa_3\left(\frac{d}{2}-p-a\right)\right].\nonumber\\
\eea

In Appendix~\ref{clp}, we show that by plugging the potential profile~(\ref{ppc2}) into Eq.~(\ref{pr1}), and using Eqs.~(\ref{t1})-(\ref{t4}), one obtains the interaction pressure in the form
\be
\label{prc1}
\beta P=\frac{\kappa_2^2-\kappa_3^2}{2\pi\ell_{\rm B} Q_+^2}.
\ee
Therein, by solving Eqs.~(\ref{t1})-({\ref{t4}) via a Taylor expansion in terms of the dipole density $\sigma\s$, we find that the pressure~(\ref{prc1}) takes the approximate closed-form
\be
\label{prc2}
\beta P\approx-2\pi\ell_{\rm B}Q_+^2\sigma\s^2\frac{\left(p+a-d/2\right)^2}{a^2}
\ee
Eq.~(\ref{prc2}) is displayed in the inset of Fig.~\ref{fig3}(b) by red dots. This limiting law indicates that as the membrane surfaces approach each other, the attractive force is amplified algebraically with the width of the dipolar overlap region. One also notes that as the dipolar docking occurs at $d=2p+a$, this attractive pressure becomes independent of the dipolar size $a$, i.e. $\beta P\approx-\pi\ell_{\rm B}Q_+^2\sigma\s^2/2$.

In order to explain the origin of the attractive force in Eq.~(\ref{prc2}), we display in Fig.~\ref{fig4}(c) the electrostatic potential profile~(\ref{ppc2}) together with the dipole configuration close to the membrane surface at $z=0$. One sees that as the dipolar layers overlap at $d<2(p+a)$,  the attraction of the dipolar cations by the electric field of the anions on the adjacent membrane wall comes into play (blue arrows). This gives rise to a second potential minimum at $z=p+a$ (black curve) that deepens with the decrease of the separation distance down to $d=2p+a$ (red curve). To determine the consequence of this energy minimum on the dipolar ordering, we substitute Eq.~(\ref{ppc2}) into the orientational probability density~(\ref{nd3}) to obtain
\bea
\label{or3}
n(a_{\rm z})&=&\frac{a\mu_-}{Q_+}\left\{\kappa_1^2\sec^2\left[\kappa_1(a_z+a)/\sqrt2\right]\theta(-a_z)\right.\\
&&\hspace{1.cm}+\kappa_2^2\sec^2\left[\kappa_2(p+a_z-z_0)/\sqrt2\right]\nonumber\\
&&\hspace{1.3cm}\times\theta(a_z)\theta(d-2p-a-a_z)\nonumber\\
&&\hspace{1.cm}+\kappa_3^2\sec^2\left[\kappa_3(p+a_z-d/2)/\sqrt2\right]\nonumber\\
&&\hspace{1.3cm}\left.\times\theta(a_z-d+2p+a)\right\}.\nonumber
\eea
Eq.~(\ref{or3}) plotted in Fig.~\ref{fig4}(d) indicates that the dipolar cation attraction towards the adjacent membrane wall translates into a second probability peak in the perpendicular dipole orientation $\theta=0$ (black curve). As the wall distance decreases further to $d=2p+a$ (red curve), this probability peak reaches the height of the second maximum at the parallel orientation $\theta=\pi/2$. 

Hence, as the membrane surfaces approach each other, the amplification of the dipolar attraction between the opposing walls  is accompanied with the reorientation of the dipoles from parallel to perpendicular configuration (see the inset of Fig.~\ref{fig4}(d)), akin to the formation of dipolar bridges across the inter-membrane space. This {\sl dipolar bridging} mechanism is precisely at the origin of the attractive membrane interactions occurring at short separation distances. It is noteworthy that such a short-ranged attractive energy component has been identified in various MD simulations of dipolar membranes~\cite{sim1,sim2}. Next, we investigate the effect of added salt on this membrane adhesion mechanism.

\subsection{Understanding the salt-mediated repulsion}

Having elucidated the mechanism governing the short-range attractive branch of the pressure curves in Fig.~\ref{fig3}(a), we investigate here the longe-range repulsive interaction regime mediated by salt. To this aim, we focus on the DH regime where the ZPB Eq.~(\ref{can14}) can be linearized and solved exactly. The derivation of this DH-level solution is presented in Appendix~\ref{ads}. Injecting the DH potential profile into Eq.~(\ref{pr1}), and Taylor-expanding the result at the quadratic order in the dipole density $\sigma\s$, the interaction pressure follows for $d\geq2p+a$ as
\bea
\label{prc4}
\frac{\beta P}{2\pi\ell_{\rm B}\sigma\s^2}&\approx&Q_{\rm eff}^2\frac{\cosh^2\left(\kas p\right)}{\sinh^2\left(\kas d/2\right)}\\
&&-Q_+^2\frac{\sinh^2\left[\kas\left(p+a-d/2\right)\right]}{\left(\kas a\right)^2}\theta\left(p+a-d/2\right),\nonumber
\eea
where we introduced the salt screening parameter $\kappa\s$ and the effective dipole charge $Q_{\rm eff}$ defined as,
\be
\label{efc}
\kappa\s=\sqrt{8\pi\ell_{\rm B}\rho_{\rm b}};\hspace{1cm}Q_{\rm eff}=Q_+\frac{\sinh\left(\kas a\right)}{\kas a}-Q_-.
\ee

Fig.~\ref{fig3}(a) shows that Eq.~(\ref{prc4}) can accurately approximate the numerical pressure curves (open circles). We now note that Eq.~(\ref{prc4}) is composed of a repulsive component (first term), and an attractive component (second term) emerging in the short distance regime $d<2(p+a)$ studied in Sec.~\ref{pa2}. Indeed, one sees that in salt-free liquids ($\kappa\s=0$), the repulsive component vanishes for $Q_-=Q_+$, and the attractive term tends to Eq.~(\ref{prc2}). 

In order to identify the origin of the repulsive term in Eq.~(\ref{prc4}), we note that its magnitude is set by the effective charge $Q_{\rm eff}$ in Eq.~(\ref{efc}). Taylor-expanding the latter, and setting $Q_-=Q_+$, one obtains the positive dipole charge $Q_{\rm eff}\approx Q_+\left(\kappa\s a\right)^2/6$  scaling linearly with the salt concentration. Thus, at large distances, the interaction of zwitterionic membranes separated by a salt solution is equivalent to the coupling of interfacially cationic walls~\cite{not3}. 

The behavior of the salt-dressed membrane dipoles as effective surface cations stems from the screening of their elementary charges. Namely, in a salt-free solution, and outside the region scanned by the rotating dipole, the field created by the dipolar cation is exactly canceled by the dipolar anion of the same magnitude. As a result, at $d>2(p+a)$, the intermembrane field and interaction pressure vanish. However, in the presence of added salt, the field of the screened dipolar anion can only partially cancel the field induced by the rotating cation. This generates a net cationic inter-membrane field giving rise to long-ranged repulsive membrane interactions.

Due to the competition between the attractive and repulsive interaction components, the net pressure~(\ref{prc4}) vanishes at the characteristic separation distance
\bea
\label{deq1}
\hspace{-2.1cm}d^*&=&p+a+\kas^{-1}{\rm arccosh}\left\{\cosh\left[\kas(p+a)\right]\right.\\
&&\hspace{3.1cm}\left.-2\kas a\cosh(\kas p)\frac{|Q_{\rm eff}|}{Q_+}\right\}.\nonumber
\eea
Setting $Q_-=Q_+$ and Taylor-expanding Eq.~(\ref{deq1}) in terms of the screening parameter, one obtains for $\kappa\s a\lesssim1$
\be
\label{deq2}
d^*\approx2(p+a)-\frac{\kappa\s a^3}{3(p+a)}.
\ee
In agreement with Fig.~\ref{fig3}(b), Eq.~(\ref{deq2}) indicates that added salt drops the equilibrium separation ($\rho_{\rm b}\uparrow d^*\downarrow$) below the close contact distance $d=2(p+a)$.

\begin{figure}
\includegraphics[width=1.0\linewidth]{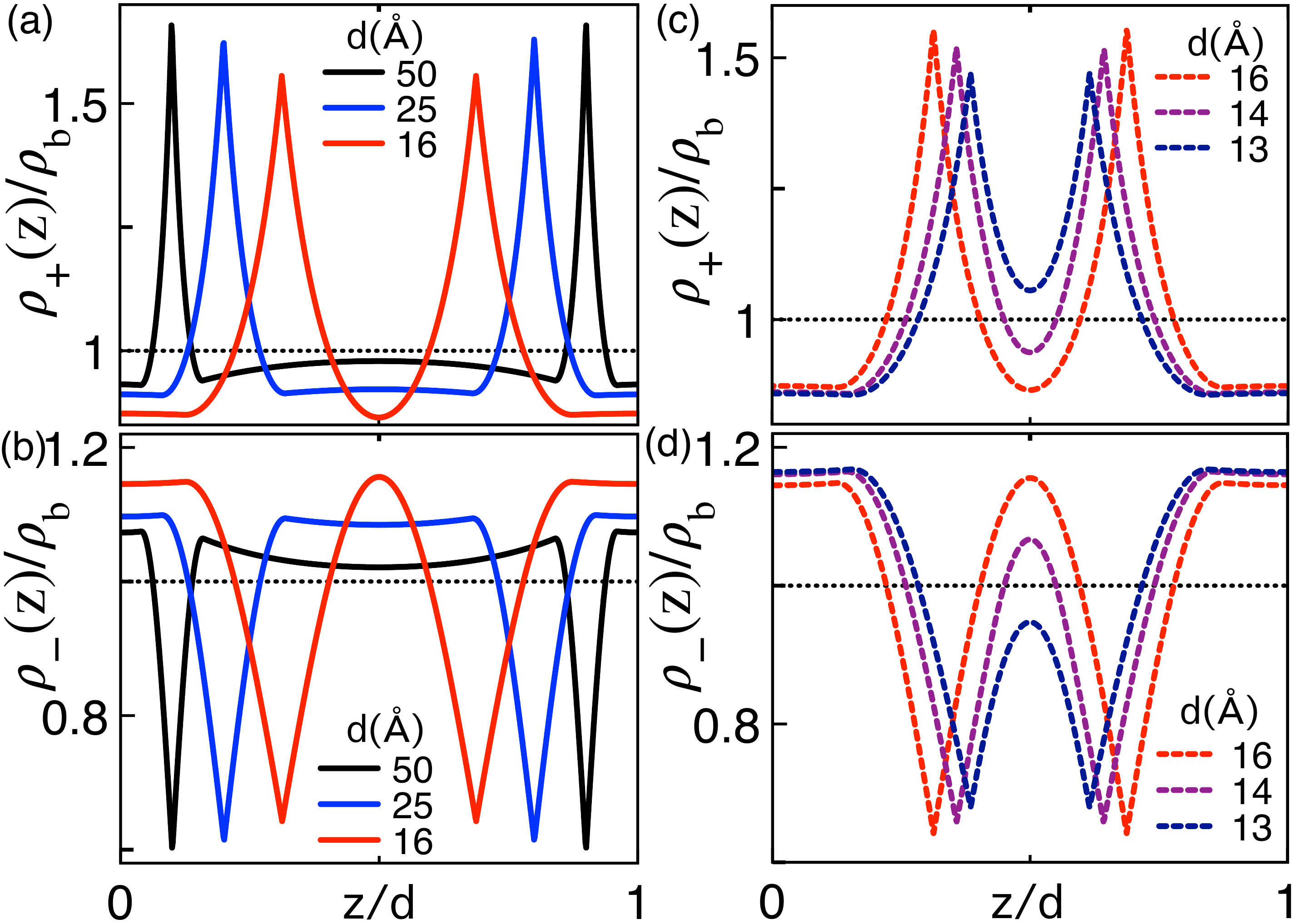}
\caption{(Color online) (a) Cation and (b) anion density profiles from Eq.~(\ref{can8}) at various pore sizes in the separation regime $d\geq2(p+a)$. (c)-(d) Density plots in (a)-(b) in the dipolar contact regime $d\leq2(p+a)$. The parameters are $p=5$ {\AA}, $a=3$ {\AA}, $Q_\pm=1$, $\rho_{\rm b}=0.1$ M, and $\sigma\s=1.0$ ${\rm e}/{\rm nm}^2$.}
\label{fig5}
\end{figure}

\subsection{Salt affinity of zwitterionic membranes and protonation effects on membrane interactions}

Figs.~\ref{fig5}(a)-(b) illustrate the salt affinity of the zwitterionic membrane in the decoupled dipole regime $d>2(p+a)$. One sees that due to the cooperative effect of the anionic and cationic fields, the salt cations (anions) are adsorbed (excluded) by the dipolar layers but excluded (adsorbed) by the intermembrane region. Moreover, as the separation distance is decreased, the amplification of the cationic intermembrane field enhances the cation depletion ($d\downarrow \rho_+(z)\downarrow$) and the anion excess ($d\downarrow \rho_-(z)\uparrow$ ) in the mid-pore area. 

Figs.~\ref{fig5}(c)-(d) show that upon the further decrease of the separation distance into the dipolar contact regime, the salt affinity of the intermembrane region is entirely reversed. For $d\leq2(p+a)$, the field of the dipolar cations is dominated by the field of the dipolar anions on the opposing wall. The resulting net anionic field responsible for the attractive dipolar bridging mechanism in Fig.~\ref{fig4}(d) switches the cation (anion) affinity of the mid-pore area from depletion (excess) to excess (exclusion) state. This intricate correlation between the slit size and the salt affinity of the dipolar membrane may have significant impacts on nano-filtration and ion transport techniques.

Previous electrophoretic transport experiments with zwitterionic lipid vesicles revealed that the regulation of the solution pH via buffer addition reduces the anionic surface charge strength of dipolar membranes~\cite{Petrache2011}. The consequence of this protonation effect on membrane interactions is qualitatively characterized in Fig.~{\ref{fig3}(c). One sees that upon the reduction of $Q_-$, the attractive interaction regime is suppressed, and the net pressure becomes more repulsive. The enhancement of the membrane repulsion by the reduction of $Q_-$ stems from the increase of the effective dipolar charge in Eq.~(\ref{efc}), and the amplification of the repulsive force component in Eq.~(\ref{prc4}), i.e. $Q_-\downarrow Q_{\rm eff}\uparrow P\uparrow$. Taylor-expanding now Eq.~(\ref{deq1}) for $Q_{\rm eff}/Q_+\lesssim1$, one obtains
\be
\label{deq3}
d^*\approx2(p+a)-2a\frac{|Q_{\rm eff}|}{Q_+}\frac{\cosh(\kappa\s p)}{\sinh\left[\kappa\s(p+a)\right]}.
\ee
In agreement with Eq.~(\ref{deq3}), Fig.~{\ref{fig3}(c) indicates that the equilibrium distance is expected to be reduced by protonation, i.e. $Q_-\downarrow Q_{\rm eff}\uparrow d^*\downarrow$.

\section{Conclusions}
\label{conc}

We developed a field theoretic model of electrolytes confined between dipolar walls. Within the framework of this model, we investigated the electrostatic interactions of zwitterionic membranes. We found that the intermembrane forces are characterized by a non-uniform behavior governed by repulsive interactions at large separation distances $d>2(p+a)$ and attractive interactions in the short distance regime $d<2(p+a)$. 

The repulsive interactions are mediated by the salt screening of the surface dipoles that causes the opposing zwitterionic walls to behave as like-charged cationic substrates. The attractive pressure branch is in turn triggered by a dipolar bridging mechanism. Namely, as the membrane walls approach each other, the dipolar cations attracted by the anions of the adjacent wall reorient themselves towards the latter to minimize their distance. This leads to the adhesion of the zwitterionic membrane walls via the electrostatic binding of their surface dipoles. 

We also observed that due to the competition between the individual field components induced by the dipolar charges, the salt affinity of the zwitterionic membrane, exhibits a complex behavior. In the large separation regime where the field of the dipolar cations governs the inter-membrane region, the latter zone is characterized by the salt anion excess and the salt cation depletion. However, the drop of the membrane separation distance into the dipolar overlap regime reverses the ion selectivity of the membrane; in the inter-membrane region, the field of the dipolar anions taking over the cationic field attracts the salt cations and excludes the salt anions. This non-uniform trend of the membrane salt affinity may have significant repercussions on water nanofiltration and nanofluidic transport techniques.

Our prediction of zwitterionic membrane attraction at short distances is partly supported by previous MD simulations where the direct interaction of the surface dipoles on the adjacent membrane walls has been indeed shown to be attractive~\cite{sim1,sim2}. However, the same works also revealed that this short-range electrostatic attraction is overpowered by repulsive hydration interactions mediated by restructuring of the solvent molecules. In order to quantify the competition between the direct electrostatic interactions and solvent-driven effects, we will generalize our model via the inclusion of explicit solvent molecules together with beyond-MF charge correlations. \SB{The inclusion of electrostatic fluctuations will also enable us to account for the additional attractive effect of thermal van der Waals interactions.} These extensions will be considered in an upcoming article.

\vspace{1cm}
{\bf Acknowledgements} 
RP acknowledges funding from the Key project \#12034019 of the National Natural Science Foundation of China.

\vspace{1cm}
{\bf Data Availability} 

The data that support the findings of this study are available from the corresponding author upon reasonable request.

\smallskip
\appendix

\section{Mapping Eq.~(\ref{can14}) to the Poisson Eq.}
\label{ap1}

Here, we show that Eq.~(\ref{can14}) can be expressed as a generalized Poisson Eq. Defining the dipolar on-site potentials as $V_{\rm l,r}(\br,\bo)= V^{-}_{\rm l,r}(\br)+V^{+}_{\rm l,r}(\br+\ba)$, the MF-level number densities of the dipolar anions $\rho^-_{\rm l,r}(\br)$ and cations $\rho^+_{\rm l,r}(\br)$ can be obtained from the thermodynamic identity
\be\label{can16}
\rho^\pm_{\rm l,r}(\br)=-\frac{\delta\ln Z_{\rm G}}{\delta V^\pm_{\rm l,r}(\br)}\approx\frac{\delta(\beta H)}{\delta V^\pm_{\rm l,r}(\br)}
\ee
as
\bea
\label{can21}
\hspace{-5mm}\rho^-_{\rm l}(z)&=&\sigma\s\delta(z-p);\hspace{5mm}\rho^-_{\rm r}(z)=\sigma\s\delta(d-z+p),\\
\label{can22}
\hspace{-5mm}\rho^+_{\rm l}(z)&=&\frac{\sigma\s}{2a I}e^{-Q_+\phi(z)}\theta_{\rm l}(z),\\
\label{can23}
\hspace{-5mm}\rho^+_{\rm r}(z)&=&\frac{\sigma\s}{2a I}e^{-Q_+\phi(z)}\theta_{\rm r}(z).
\eea
By identifying now the number densities~(\ref{can21})-(\ref{can23}) in the ZPB Eq.~(\ref{can14}), one can recast the latter as a Poisson Eq.,
\be
\label{can24}
\frac{k_{\rm B}T}{e^2}\partial_z\e(z)\partial_z\phi(z)+\sum_{n=1}^sq_n\rho_n(z)+\sum_{\alpha={\rm l,r}}\sum_{\gamma=\pm}Q_\gamma\rho^\gamma_\alpha(z)=0.
\ee

\section{Calculating the interaction pressure from the first integral of the ZPB Eq.~(\ref{can14})}
\label{apcv}

In this appendix, we calculate the first integral of the ZPB Eq.~(\ref{can14}) and derive a general form for the zwitterionic membrane pressure. To this aim, we first multiply Eq.~(\ref{can14}) with the local electric field $\phi'(z)$. This yields
\bea
\label{apcv1}
&&\phi'(z)\phi''(z)+4\pi\ell_{\rm B}\sum_{n=1}^s\rho_{{\rm b}n}q_n\phi'(z)e^{-q_n\phi(z)}\nonumber\\
&&-4\pi\ell_{\rm B}\sigma\s Q_-\phi'(z)\left[\delta(z-p)+\delta(z-d+p)\right]\\
&&+\frac{2\pi\ell_{\rm B}}{aI}\sigma\s Q_+\phi'(z)e^{-Q_+\phi(z)}\left[\theta_{\rm l}(z)+\theta_{\rm r}(z)\right]=0.\nonumber
\eea
Next, we express Eq.~(\ref{apcv1}) as a total derivative, i.e.
\bea\label{apcv2}
&&-\frac{1}{8\pi\ell_{\rm B}}\frac{{\rm d}}{dz}\left(\frac{{\rm d}\phi}{{\rm d}z}\right)^2+\frac{{\rm d}}{dz}\sum_{n=1}^s\rho_{{\rm b}n}e^{-q_n\phi(z)}\nonumber\\
&&+\sigma\s Q_-\frac{{\rm d}F(z)}{dz}-\frac{\sigma\s}{2aI}\frac{{\rm d}G(z)}{dz}=0,
\eea
where the functions $F(z)$ and $G(z)$ satisfy the Eqs.
\bea
\label{apcv4}
&&\frac{{\rm d}F(z)}{{\rm d}z}=\phi'(z)\left[\delta(z-p)+\delta(z-d+p)\right],\\
\label{apcv3}
&&\frac{{\rm d}G(z)}{{\rm d}z}=Q_+\phi'(z)e^{-Q_+\phi(z)}\left[\theta_{\rm l}(z)+\theta_{\rm r}(z)\right].
\eea
Upon the formal integration of Eq.~(\ref{apcv2}) from $z'=0$ to $z'=z$, the first integral $C(z)$ of Eq.~(\ref{apcv1}) becomes
\be
\label{apcv5}
C(z)=-\frac{\phi'(z)^2}{8\pi\ell_{\rm B}}+\sum_{n=1}^s\left[\rho_i(z)-\rho_i(0)\right]+\sigma\s Q_-F(z)-\frac{\sigma\s G(z)}{2aI},
\ee
where the functions $F(z)$ and $G(z)$ follow from Eqs.~(\ref{apcv3})-(\ref{apcv4}) as
\bea
\label{apcv7}
&&\hspace{-7mm}F(z)=\int_0^z{\rm d}z'\phi'(z')\left[\delta(z'-p)+\delta(z'-d+p)\right],\\
\label{apcv6}
&&\hspace{-7mm}G(z)=Q_+\int_0^z{\rm d}z'\phi'(z')e^{-Q_+\phi(z')}\left[\theta_{\rm l}(z')+\theta_{\rm r}(z')\right].
\eea

Taking into account the discontinuity of the field $\phi'(z)$ at $z=p$ and $z=d-p$, the evaluation of the integral in Eq.~(\ref{apcv7}) yields
\bea
\label{apcv8}
F(z)&=&\frac{1}{2}\left[\phi'(p^+)+\phi'(p^-)\right]\theta(z-p)\\
&&+\frac{1}{2}\left\{\phi'\left[(d-p)^+\right]+\phi'\left[(d-p)^-\right]\right\}\theta(z-d+p).\nonumber
\eea
Then, in order to obtain the function $G(z)$ from the integral~(\ref{apcv6}), with the theta functions defined in Eqs.~(\ref{tl1})-(\ref{tl2}), we carry out an integration by parts, and use the identity $\partial_x\theta(x)=\delta(x)$. After some algebra, one obtains
\bea
\label{apcv9}
G(z)&=&-e^{-Q_+\phi(z)}\left[\theta_{\rm l}(z)+\theta_{\rm r}(z)\right]\\
&&+e^{-Q_+\phi(p-a)}\theta(z-p+a)\nonumber\\
&&-e^{-Q_+\phi(p+a)}\theta(z-p-a)\nonumber\\
&&+e^{-Q_+\phi(d-p-a)}\theta(z-d+p+a)\nonumber\\
&&-e^{-Q_+\phi(d-p+a)}\theta(z-d+p-a).\nonumber
\eea
Finally, plugging Eqs.~(\ref{apcv8})-(\ref{apcv9}) into Eq.~(\ref{apcv5}), the first integral of Eq.~(\ref{can14}) takes the form
\bea
\label{apcv10}
C(z)&=&-\frac{\phi'(z)^2}{8\pi\ell_{\rm B}}+\sum_{n=1}^s\left[\rho_n(z)-\rho_n(0)\right]\\
&&+\frac{\sigma\s}{2aI}e^{-Q_+\phi(z)}\left[\theta_{\rm l}(z)+\theta_{\rm r}(z)\right]\nonumber\\
&&-\frac{\sigma\s}{2aI}\left\{e^{-Q_+\phi(p-a)}\theta(z-p+a)\right.\nonumber\\
&&\hspace{1.2cm}-e^{-Q_+\phi(p+a)}\theta(z-p-a)\nonumber\\
&&\hspace{1.2cm}+e^{-Q_+\phi(d-p-a)}\theta(z-d+p+a)\nonumber\\
&&\hspace{1.2cm}\left.-e^{-Q_+\phi(d-p+a)}\theta(z-d+p-a)\right\}\nonumber\\
&&+\frac{\sigma\s Q_-}{2}\left[\phi'(p^+)+\phi'(p^-)\right]\theta(z-p)\nonumber\\
&&+\frac{\sigma\s Q_-}{2}\left\{\phi'\left[(d-p)^+\right]+\phi'\left[(d-p)^-\right]\right\}\nonumber\\
&&\hspace{1.25cm}\times\theta(z-d+p).\nonumber
\eea

In order to derive the net pressure from the first integral~(\ref{apcv10}), we will normalize the latter with its value in the infinite membrane separation regime. To this aim, we set $z=d/2$ and take the limit $d\to\infty$. This yields
\bea
\label{apcv11}
C_{\rm b}&=&\sum_{n=1}^s\left[\rho_{n{\rm b}}-\rho_n(0)\right]+\frac{\sigma\s Q_-}{2}\left[\phi'(p^+)+\phi'(p^-)\right]\nonumber\\
&&+\frac{\sigma\s}{2aI}\left\{e^{-Q_+\phi(p+a)}-e^{-Q_+\phi(p-a)}\right\}.
\eea
Subtracting from Eq.~(\ref{apcv10}) the constant~(\ref{apcv11}), and using the dipolar density functions~(\ref{pr2}) together with the identity $\theta(z-z')-1=-\theta(z'-z)$, the normalized first integral corresponding to the net interplate pressure $\beta P=C(z)- C_{\rm b}$ follows as
\bea
\label{apcv12}
\beta P&=&-\frac{\phi'(z)^2}{8\pi\ell_{\rm B}}+\sum_{n=1}^s\left[\rho_n(z)-\rho_{n{\rm b}}\right]+\rho^+_{\rm r}(z)+\rho^+_{\rm l}(z)\nonumber\\
&&+\rho^+_{\rm r}(d-p+a)\theta(z-d+p-a)\nonumber\\
&&-\rho^+_{\rm r}(d-p-a)\theta(z-d+p+a)\nonumber\\
&&+\rho^+_{\rm l}(p-a)\theta(p-a-z)-\rho^+_{\rm l}(p+a)\theta(p+a-z)\nonumber\\
&&+\frac{\sigma\s Q_-}{2}\left\{\phi'\left[(d-p)^+\right]+\phi'\left[(d-p)^-\right]\right\}\nonumber\\
&&\hspace{1.25cm}\times\theta(z-d+p)\nonumber\\
&&-\frac{\sigma\s Q_-}{2}\left[\phi'(p^+)+\phi'(p^-)\right]\theta(p-z).
\eea
We also note that as the mirror symmetry $\phi(d-z)=\phi(z)$ implies $\rho^+_{\rm l}(p\pm a)=\rho^+_{\rm r}(d-p\mp a)$ and $\phi'(p^\pm)=\mp\phi'\left[(d-p)^\mp\right]$, the pressure~(\ref{apcv12}) can be further simplified as
\bea
\label{apcv13}
\beta P&=&-\frac{\phi'(z)^2}{8\pi\ell_{\rm B}}+\sum_{n=1}^s\left[\rho_n(z)-\rho_{n{\rm b}}\right]+\rho^+_{\rm r}(z)+\rho^+_{\rm l}(z)\nonumber\\
&&+\rho^+_{\rm r}(d-p+a)\left[\theta(z-d+p-a)+\theta(p-a-z)\right]\nonumber\\
&&-\rho^+_{\rm r}(d-p-a)\left[\theta(z-d+p+a)+\theta(p+a-z)\right]\nonumber\\
&&+\frac{\sigma\s Q_-}{2}\left\{\phi'\left[(d-p)^+\right]+\phi'\left[(d-p)^-\right]\right\}\\
&&\hspace{1.25cm}\times\left[\theta(z-d+p)+\theta(p-z)\right].\nonumber
\eea

In agreement with the mechanical equilibrium condition, the r.h.s. of the first integral~(\ref{apcv12}) is naturally independent of the coordinate $z$. In Sec.~\ref{cn}, we use Eq.~(\ref{apcv12}) to relate the interaction pressure to the charge density values in different zones of the intermembrane region.

\section{Deriving the orientational probability~(\ref{nd3})}
\label{ap1II}

We derive here the orientational probability characterizing the rotational configuration of the surface dipoles. To this aim, we redefine the dipolar on-site potentials as $V_{\rm l,r}(\br,\bo)\to V_{\rm l,r}(\br)$ such that they act only on the dipolar anions fixed at $\br$. This allows to obtain the fixed anion number densities as
\be
\label{nd1}
\rho^-_\alpha(\br)=\left.\frac{\delta(\beta H)}{\delta V^\pm_\alpha(\br)}\right|_{V_\alpha=0}=:\sigma\s\delta(z-z\ce)\int\frac{\mathrm{d}^2\bo}{4\pi}n_\alpha(\bo),
\ee
with the auxiliary parameter $z\ce=p$ if $\alpha=l$, and $z\ce=d-p$ for $\alpha=r$. The second equality of Eq.~(\ref{nd1}) defines the orientational probability such that the latter yields the correct dipolar anion density $\rho^-_\alpha(z)=\sigma\s\delta(z-z\ce)$ and normalization condition $\int\mathrm{d}^2\bo\;n_\alpha(\bo)/(4\pi)=1$. Evaluating the derivative in Eq.~(\ref{nd1}), and accounting for the symmetry with respect to the mid-pore, the orientational probabilities follow as $n_{\rm l}(a_{\rm z})=n_{\rm r}(a_{\rm z})=n(a_{\rm z})$, with
\be
\label{and3}
n(a_{\rm z})=\frac{1}{I}e^{-Q_+\phi(p+a_z)}.
\ee

\section{Salt-free solution of the ZPB Eq.~(\ref{can14})}
\label{ap2}

In this Appendix, we derive the exact solution of the ZPB Eq.~(\ref{can14}) in a salt-free solution. Setting $\rho_{n{\rm b}}=0$, Eq.~(\ref{can14}) becomes
\bea
\label{can25}
&&\frac{k_{\rm B}T}{e^2}\partial_z\e(z)\partial_z\phi(z)-\sigma\s Q_-\left[\delta(z-p)+\delta(z-d+p)\right]\nonumber\\
&&+\frac{\sigma\s Q_+}{2aI}\left\{e^{-Q_+\phi(z)}\theta_{\rm l}(z)+e^{-Q_+\phi(z)}\theta_{\rm r}(z)\right\}=0.
\eea
We now note that according to Eqs.~(\ref{tl1})-(\ref{tl2}), the rotating dipole densities in the curly bracket term of Eq.~(\ref{can25}) overlap for $d\leq2(p+a)$. This implies that Eq.~(\ref{can25}) should be solved in two different distance regimes. 

\subsection{Derivation of the solution~(\ref{ppc}) for $d\geq2(p+a)$}

We first consider the membrane separation regime $d>2(p+a)$ where the surface dipoles on different walls do not overlap. Due to the mirror symmetry with respect to the point $z=d/2$, and the theta functions~(\ref{tl1})-(\ref{tl2}) fixing the regions scanned by the rotating dipoles, the solution to Eq.~(\ref{can25}) has the piecewise form
\bea
\label{can26}
\phi(z)&=&\phi_0\theta(z)\theta(p-a-z)+\phi_1(z)\theta(z-p+a)\theta(p-z)\nonumber\\
&&+\phi_2(z)\theta(z-p)\theta(p+a-z)\nonumber\\
&&+\phi_3\theta(z-p-a)\theta(d/2-z)\\
&&+\phi(d-z)\theta(z-d/2)\theta(d-z).\nonumber
\eea
In Eq.~(\ref{can26}), $\phi_0$ and $\phi_3$ are constant potentials in the unscreened slit zones inaccessible to the dipolar cations. The remaining potential components with index $i=1,2$ solve the differential equation
\be
\label{can27}
\frac{d^2\phi_i(z)}{dz^2}+\frac{1}{a\mu_+I}e^{-Q_+\phi_i(z)}=0.
\ee

Multiplying Eq.~(\ref{can27}) by $d\phi_i(z)/dz$, integrating the result, and imposing the continuity of the potential and the field at $z=p\pm a$, the first integrals of Eq.~(\ref{can27}) follow as
\bea
\label{can28}
\frac{d\phi_1(z)}{dz}&=&\frac{\sqrt2\kappa_0}{Q_+}\left\{e^{-Q_+\left[\phi_1(z)-\phi_0\right]}-1\right\}^{1/2},\\
\label{can28II}
\frac{d\phi_2(z)}{dz}&=&\frac{\sqrt2\kappa'_0}{Q_+}\left\{e^{-Q_+\left[\phi_2(z)-\phi_3\right]}-1\right\}^{1/2},
\eea
where $\kappa_0$ and $\kappa'_0$ are integration constants. Solving Eqs.~(\ref{can28})-(\ref{can28II}), and imposing at $z=p$ the continuity of the potential together with the Gauss' law in Eq.~(\ref{bc1}), one finally obtains $\phi_3=\phi_0$, $\kappa'_0=\kappa_0$ with
\be\label{ts1A}
\frac{\kappa_0 a}{\sqrt2}\tan\left(\frac{\kappa_0 a}{\sqrt2}\right)=\frac{Q_+a}{2\mu_-},
\ee
and the non-uniform potential components in Eq.~(\ref{can26}),
\bea
\label{c2A}
&&\hspace{-5mm}\phi_1(z)=\phi_0+\frac{1}{Q_+}\ln\left\{\cos^2\left[\frac{\kappa_0}{\sqrt 2}(z-p+a)\right]\right\},\\
\label{c2IIA}
&&\hspace{-5mm}\phi_2(z)=\phi_0+\frac{1}{Q_+}\ln\left\{\cos^2\left[\frac{\kappa_0}{\sqrt 2}(z-p-a)\right]\right\}.
\eea

\subsection{Derivation of the solution~(\ref{ppc2}) for $2(p+a)\geq d\geq2p+a$}
\label{apx}

\subsubsection{Deriving the piecewise potential}

In the slit size regime $2(p+a)\geq d\geq2p+a$, the rotating dipoles on the left and right walls overlap at $d-p-a\leq z\leq p+a$. As a result, the mid-pore region is characterized by a non-uniform potential. Thus, the piecewise solution to Eq.~(\ref{can25}) reads
\bea
\label{can29}
\phi(z)&=&\phi_0\theta(z)\theta(p-a-z)+\phi_1(z)\theta(z-p+a)\theta(p-z)\nonumber\\
&&+\phi_2(z)\theta(z-p)\theta(d-p-a-z)\nonumber\\
&&+\phi_3(z)\theta(z-d+p+a)\theta(d/2-z)\nonumber\\
&&+\phi(d-z)\theta(z-d/2)\theta(d-z).
\eea
In Eq.~(\ref{can29}), $\phi_0$ is the uniform potential component in the pure solvent region. Then, the non-uniform potential components solve the differential Eqs.
\be
\label{can30}
\frac{d^2\phi_i(z)}{dz^2}+\frac{1}{a\mu_+I}e^{-Q_+\phi_i(z)}=0
\ee
for $i=1,2$, and
\be
\label{can31}
\frac{d^2\phi_3(z)}{dz^2}+\frac{2}{a\mu_+I}e^{-Q_+\phi_3(z)}=0
\ee
in the dipolar overlap region. 

Integrating Eqs.~(\ref{can30})-(\ref{can31}), and imposing the continuity of the solution through the slit, the first integrals follow as
\bea
\label{can32}
\hspace{-5mm}\frac{d\phi_1(z)}{dz}&=&\frac{\sqrt2\kappa_1}{Q_+}\left\{e^{-Q_+\left[\phi_1(z)-\phi_0\right]}-1\right\}^{1/2},\\
\label{can33}
\hspace{-5mm}\frac{d\phi_2(z)}{dz}&=&\frac{\sqrt2\kappa_2}{Q_+}\left\{\frac{\kappa_1^2}{\kappa_2^2}e^{-Q_+\left[\phi_2(z)-\phi_0\right]}-1\right\}^{1/2},\\
\label{can34}
\hspace{-5mm}\frac{d\phi_3(z)}{dz}&=&\frac{2\kappa_3}{Q_+}\left\{\frac{\kappa_1^2}{\kappa_3^2}e^{-Q_+\left[\phi_3(z)-\phi_0\right]}-1\right\}^{1/2},
\eea
where $\kappa_i$ with $i=\{1,2,3\}$ are integration constants. Solving Eqs.~(\ref{can32})-(\ref{can34}), and imposing the continuity of the field and potential together with the Gauss' law in Eq.~(\ref{bc1}), the potential components finally follow as
\bea
\label{can35}
&&\hspace{-8mm}\phi_1(z)=\phi_0+\frac{1}{Q_+}\ln\left\{\cos^2\left[\frac{\kappa_1}{\sqrt2}(z-p+a)\right]\right\},\\
\label{can36}
&&\hspace{-8mm}\phi_2(z)=\phi_0+\frac{1}{Q_+}\ln\left\{\frac{\kappa_1^2}{\kappa_2^2}\cos^2\left[\frac{\kappa_2}{\sqrt2}(z-z_0)\right]\right\},\\
\label{can37}
&&\hspace{-8mm}\phi_3(z)=\phi_0+\frac{1}{Q_+}\ln\left\{\frac{\kappa_1^2}{\kappa_3^2}\cos^2\left[\kappa_3\left(z-\frac{d}{2}\right)\right]\right\},
\eea
where the constants $\kappa_i$ and $z_0$ satisfy the non-linear Eqs.
\bea\label{can38}
&&\hspace{-3mm}\cos\left[\frac{\kappa_2}{\sqrt2}(p-z_0)\right]=\frac{\kappa_2}{\kappa_1}\cos\left(\frac{\kappa_1a}{\sqrt2}\right),\\
\label{can39}
&&\hspace{-3mm}\tan\left[\frac{\kappa_2}{\sqrt2}(p-z_0)\right]=\frac{\kappa_1}{\kappa_2}\tan\left(\frac{\kappa_1a}{\sqrt2}\right)-\frac{\sqrt2Q_+}{\kappa_2\mu_-},\\
\label{can40}
&&\hspace{-3mm}\cos\left[\frac{\kappa_2}{\sqrt2}(d-p-a-z_0)\right]=\frac{\kappa_2}{\kappa_3}\cos\left[\kappa_3\left(\frac{d}{2}-p-a\right)\right],\nonumber\\
&&\\
\label{can41}
&&\hspace{-3mm}\sin\left[\frac{\kappa_2}{\sqrt2}(d-p-a-z_0)\right]=\sqrt2\sin\left[\kappa_3\left(\frac{d}{2}-p-a\right)\right].\nonumber\\
\eea
We finally note that if one evaluates the integral $I$ in Eq.~(\ref{can15}) with the salt-free potential solutions~(\ref{can26}) and~(\ref{can29}) derived above, one obtains the self-consistency condition $Q_-=Q_+$. The latter equality corresponds to the interfacial electroneutrality constraint originating from the absence of salt.

\subsubsection{Deriving the analytical Eqs.~(\ref{prc1}) and~(\ref{prc2}) for pressure}
\label{clp}

Here, we derive closed form expressions for the intermembrane pressure in the salt-free liquid. To this aim, we substitute the potential profile~(\ref{can29}) into Eq.~(\ref{pr1}). This yields
\bea
\label{pra1}
\beta P&=&\frac{\mu_+\sigma\s\kappa_1^2}{2Q_+}\left\{1-\frac{\kappa_3^2}{\kappa_1^2}\sec^2\left[\kappa_3\left(\frac{d}{2}-p-a\right)\right]\right\}\nonumber\\
&&+\sigma\s\left\{\sqrt2\kappa_1\tan\left(\frac{\kappa_1a}{\sqrt2}\right)-\frac{Q_+}{\mu_-}\right\}.
\eea
Next, we combine Eqs.~(\ref{can40}) and~(\ref{can41}) to get
\be
\label{pra2}
\kappa_3^2\sec^2\left[\kappa_3\left(\frac{d}{2}-p-a\right)\right]=2\kappa_3^2-\kappa_2^2.
\ee
Then, combining Eqs.~(\ref{can38}) and~(\ref{can39}), one obtains
\be\label{pra3}
\sqrt2\kappa_1\tan\left(\frac{\kappa_1a}{\sqrt2}\right)-\frac{Q_+}{\mu_-}=\frac{\kappa_2^2\mu_-}{\sqrt2Q_+}\left(1-\frac{\kappa_1^2}{\kappa_2^2}\right).
\ee
Plugging the identities~(\ref{pra2})-(\ref{pra3}) into Eq.~(\ref{pra1}), the interplate pressure finally simplifies to
\be
\label{pra4}
\beta P=\frac{\kappa_2^2-\kappa_3^2}{2\pi\ell_{\rm B} Q_+^2}.
\ee

In order to obtain an insightful expression for the pressure~(\ref{pra4}), we carry-out a perturbative expansion of the transcendental Eqs.~(\ref{can38})-(\ref{can41}) in terms of the dipole density $\sigma\s$. To this aim, we inject into these equations the perturbative series $z_0\approx z_{01}+z_{02}\sqrt{\sigma\s}+z_{03}\sigma\s+z_{04}\sigma\s^{3/2}$, and $\kappa_i\approx\kappa_{i1}\sqrt{\sigma\s}+\kappa_{i2}\sigma\s+\kappa_{i3}\sigma\s^{3/2}$ for $1\leq i\leq3$. This yields
\bea
\label{eqp1}
&&z_0\approx p+a,\\
\label{eqp2}
&&\kappa_1\approx\kappa_2\approx\sqrt{\frac{Q_+}{a\mu_-}}-\frac{a^2}{12}\left(\frac{Q_+}{a\mu_-}\right)^{3/2},\\
\label{eqp3}
&&\kappa_3\approx\sqrt{\frac{Q_+}{a\mu_-}}+\frac{1}{24}\left(\frac{Q_+}{a\mu_-}\right)^{3/2}\left[3(2p+2a-d)^2-2a^2\right].\nonumber\\
\eea
Plugging the approximative solutions~(\ref{eqp2})-(\ref{eqp3}) into Eq.~(\ref{pra4}), the dipolar membrane pressure follows as
\be
\label{pra5}
\beta P\approx-2\pi\ell_{\rm B}Q_+^2\sigma\s^2\frac{\left(p+a-d/2\right)^2}{a^2}.
\ee

\section{DH-level solution of the ZPB Eq.~(\ref{can14}) with added salt}
\label{ads}

In this appendix, we derive the DH-level analytical solution of the ZPB Eq.~(\ref{can14}) for a $1:1$ salt solution. Thus, we set $q_\pm=\pm1$ and $\rho_{{\rm b}n}=\rho_{\rm b}$, and linearize Eq.~(\ref{can14}) in terms of the potential $\phi(z)$. This yields
\bea
\label{lzpb}
&&\partial_z^2\phi(z)-\left\{\kappa\s^2+(\xi_1^2-\kappa\s^2)\left[\theta_{\rm l}(z)+\theta_{\rm r}(z)\right]\right\}\phi(z)\\
&&=\frac{2}{\mu_-}\left[\delta(z-p)+\delta(z-d+p)\right]-\psi\left[\theta_{\rm l}(z)+\theta_{\rm r}(z)\right],\nonumber
\eea
where we defined the auxiliary parameters
\be
\label{sc}
\kappa\s^2=8\pi\ell_{\rm B}\rho_{\rm b};\hspace{5mm}\xi_1^2=\kappa\s^2+Q_+\psi;\hspace{5mm}\psi=\frac{1}{\mu_+aI}.
\ee

\subsection{Solution of Eq.~(\ref{lzpb}) for $d\geq2(p+a)$}
\label{pas1}

In the separation regime $d\geq2(p+a)$, the potential solution to Eq.~(\ref{lzpb}) follows in the piecewise form 
\bea
\label{sppc}
\phi(z)&=&\phi_1(z)\theta(z)\theta(p-a-z)+\phi_2(z)\theta(z-p+a)\theta(p-z)\nonumber\\
&&+\phi_3(z)\theta(z-p)\theta(p+a-z)\nonumber\\
&&+\phi_4(z)\theta(z-p-a)\theta(d/2-z)\\
&&+\phi(d-z)\theta(z-d/2)\theta(d-z),\nonumber
\eea
with the potential components
\bea
\label{p1}
\phi_1(z)&=&b_1\cosh(\kas z),\\
\label{p1}
\phi_2(z)&=&b_2e^{\xi_1 z}+b_3 e^{-\xi_1 z}+\frac{\psi}{\xi_1^2},\\
\label{p3}
\phi_3(z)&=&b_4e^{\xi_1 z}+b_5 e^{-\xi_1 z}+\frac{\psi}{\xi_1^2},\\
\label{p4}
\phi_4(z)&=&b_6\cosh\left[\kas\left(\frac{d}{2}-z\right)\right].
\eea
Imposing the continuity of the potential~(\ref{sppc}) throughout the intermembrane region together with the Gauss' law~(\ref{bc1}), the coefficients in Eqs.~(\ref{p1})-(\ref{p4}) follow as
\bea
\label{co1}
b_1&=&\frac{1}{\gamma_1\sinh\left[\kas(p-a)\right]}\left\{(1-\Delta_1)e^{\xi_1(p-a)}b_2+\frac{\Delta_0\psi}{\xi_1^2}\right\},\nonumber\\
&&\\
\label{co2}
b_2&=&-\frac{1}{s_-}\frac{1+\Delta_3e^{-2\xi_1a}}{1-\Delta_1\Delta_3 e^{-4\xi_1a}}e^{-\xi_1p}\\
&&-\frac{\psi}{\xi_1^2}\frac{\Delta_2+\Delta_0\Delta_3e^{-2\xi_1a}}{1-\Delta_1\Delta_3e^{-4\xi_1a}}e^{-\xi_1(p+a)},\nonumber\\
\label{co3}
b_3&=&b_2\Delta_1e^{2\xi_1(p-a)}-\frac{\Delta_0\psi}{\xi_1^2}e^{\xi_1(p-a)};\;b_4=b_2+\frac{e^{-\xi_1p}}{s_-},\nonumber\\
&&\\
\label{co4}
b_5&=&b_2\Delta_1e^{2\xi_1(p-a)}-\frac{\Delta_0\psi}{\xi_1^2}e^{\xi_1(p-a)}-\frac{e^{\xi_1p}}{s_-},\\
\label{co5}
b_6&=&-\frac{1}{\gamma_1\sinh\left[\kas\left(\frac{d}{2}-p-a\right)\right]}\\
&&\hspace{3mm}\times\left\{b_2\left(1-\Delta_1e^{-4\xi_1a}\right)e^{\xi_1(p+a)}\right.\nonumber\\
&&\hspace{9mm}\left.+\frac{2}{s_-}\cosh(\xi_1a)+\frac{\Delta_0\psi}{\xi_1^2}e^{-2\xi_1a}\right\},\nonumber
\eea
where we defined the auxiliary parameters $s_-=\xi_1\mu_-$, $\gamma_1=\kappa\s/\xi_1$, and
\bea
\label{par1}
\hspace{-1cm}\Delta_0&=&\frac{\gamma_1\tanh\left[\kas(p-a)\right]}{1+\gamma_1\tanh\left[\kas(p-a)\right]}; \;\Delta_1=1-2\Delta_0,\\
\hspace{-1cm}\Delta_2&=&\frac{\gamma_1\tanh\left[\kas\left(\frac{d}{2}-p-a\right)\right]}{1+\gamma_1\tanh\left[\kas\left(\frac{d}{2}-p-a\right)\right]},\\
\hspace{-1cm}\Delta_3&=&1-2\Delta_2.
\eea

\subsection{Solution of Eq.~(\ref{lzpb}) for $2(p+a)\geq d\geq2p+a$}
\label{pas2}

In the distance regime $2(p+a)\geq d\geq2p+a$, the piecewise potential solving Eq.~(\ref{lzpb}) has the form 
\bea
\label{sppc2}
\phi(z)&=&\phi_1(z)\theta(z)\theta(p-a-z)\\
&&+\phi_2(z)\theta(z-p+a)\theta(p-z)\nonumber\\
&&+\phi_3(z)\theta(z-p)\theta(d-p-a-z)\nonumber\\
&&+\phi_4(z)\theta(z-d+p+a)\theta(d/2-z)\nonumber\\
&&+\phi(d-z)\theta(z-d/2)\theta(d-z),\nonumber
\eea
with the potential components
\bea
\label{p5}
\phi_1(z)&=&c_1\cosh(\kas z),\\
\label{p6}
\phi_2(z)&=&c_2e^{\xi_1 z}+c_3 e^{-\xi_1 z}+\frac{\psi}{\xi_1^2},\\
\label{p7}
\phi_3(z)&=&c_4e^{\xi_1 z}+c_5 e^{-\xi_1 z}+\frac{\psi}{\xi_1^2},\\
\label{p8}
\phi_4(z)&=&c_6\cosh\left[\xi_2\left(\frac{d}{2}-z\right)\right]+\frac{2\psi}{\xi_2^2}.
\eea
In Eqs.~(\ref{p5})-(\ref{p8}), we defined the additional parameters $\xi_2^2=\kappa\s^2+2Q_+\psi$ and $\gamma_2=\xi_2/\xi_1$, and the coefficients
\bea
\label{co1}
c_1&=&\frac{1}{\gamma_1\sinh\left[\kas(p-a)\right]}\left\{(1-\Delta_1)e^{\xi_1(p-a)}c_2+\frac{\Delta_0\psi}{\xi_1^2}\right\},\nonumber\\
&&\\
\label{co2}
c_2&=&-\frac{1}{s_-}\frac{e^{\xi_1(p+2a)}+\Delta_5e^{\xi_1(2d-3p)}}{\Delta_5e^{2\xi_1(d-p)}-\Delta_1e^{2\xi_1p}}\\
&&+\frac{\psi}{\Delta_5e^{2\xi_1(d-p)}-\Delta_1e^{2\xi_1p}}\nonumber\\
&&\hspace{3mm}\times\left\{\frac{1}{\xi_1^2}\left[\Delta_4e^{\xi_1(d-p+a)}-\Delta_0e^{\xi_1(p+a)}\right]\right.\nonumber\\
&&\hspace{9mm}\left.-\frac{2}{\xi_2^2}\Delta_4e^{\xi_1(d-p+a)}\right\},\nonumber\\
c_3&=&c_2e^{2\xi_1(p-a)}\Delta_1-\frac{\psi}{\xi_1^2}\Delta_0e^{\xi_1(p-a)};\;c_4=c_2+\frac{e^{-\xi_1p}}{s_-},\nonumber\\
&&\\
c_5&=&\Delta_5\left(c_2+\frac{1}{s_-}e^{-\xi_1p}\right)e^{2\xi_1(d-p-a)}\\
&&-\Delta_4\psi\left(\frac{1}{\xi_1^2}-\frac{2}{\xi_2^2}\right)e^{\xi_1(d-p-a)},\nonumber\\
c_6&=&\frac{1}{\gamma_2\sinh\left[\xi_2\left(\frac{d}{2}-p-a\right)\right]}\\
&&\times\left\{(1-\Delta_5)\left(c_2+\frac{e^{-\xi_1p}}{s_-}\right)e^{\xi_1(d-p-a)}\right.\nonumber\\
&&\hspace{5mm}\left.+\Delta_4\psi\left(\frac{1}{\xi_1^2}-\frac{2}{\xi_2^2}\right)\right\},\nonumber
\eea
with
\be
\label{par2}
\Delta_4=\frac{\gamma_2\tanh\left[\xi_2\left(\frac{d}{2}-p-a\right)\right]}{1+\gamma_2\tanh\left[\xi_2\left(\frac{d}{2}-p-a\right)\right]}; \;\Delta_5=1-2\Delta_4.
\ee


\begin{thebibliography}{99}
\bibitem {not3} We note that a similar effect has been previously observed in membrane models with fixed interfacial dipoles~\cite{DipFix}.
\end{thebibliography}
\end{document}